# Millisecond single-molecule localization microscopy combined with convolution analysis and automated image segmentation to determine protein concentrations in complexly structured, functional cells, one cell at a time

*Adam J. M. Wollman, Mark C. Leake*

*Biological Physicsal Sciences Institute (BPSI)*
*Department of Physics and Biology*
*University of York*
*York YO10 5DD, UK.*

*T: +44 (0) 1904 322697*
*mark.leake@york.ac.uk*

**Abstract**

We present a single-molecule tool called the CoPro (Concentration of Proteins) method that uses millisecond imaging with convolution analysis, automated image segmentation and super-resolution localization microscopy to generate robust estimates for protein concentration in different compartments of single living cells, validated using realistic simulations of complex multiple compartment cell types. We demonstrates its utility experimentally on model *Escherichia coli* bacteria and *Saccharomyces cerevisiae* budding yeast cells, and use it to address the biological question of how signals are transduced in cells. Cells in all domains of life dynamically sense their environment through signal transduction mechanisms, many involving gene regulation. The glucose sensing mechanism of *S. cerevisiae* is a model system for studying gene regulatory signal transduction. It uses the multi-copy expression inhibitor of the GAL gene family, Mig1, to repress unwanted genes in the presence of elevated extracellular glucose concentrations. We fluorescently labelled Mig1 molecules with green fluorescent protein (GFP) via chromosomal integration at physiological expression levels in living *S. cerevisiae* cells, in addition to the RNA polymerase protein Nrd1 with the fluorescent protein reporter mCherry. Using CoPro we make quantitative estimates of Mig1 and Nrd1 protein concentrations in the cytoplasm and nucleus compartments on a cell-by-cell basis under physiological conditions.  These estimates indicate a ~4-fold shift towards higher values in concentration of diffusive Mig1 in the nucleus if the external glucose concentration is raised, whereas equivalent levels in the cytoplasm shift to smaller values with a relative change an order of magnitude smaller. This compares with Nrd1 which is not involved directly in glucose sensing, which is almost exclusively localized in the nucleus under high and low external glucose levels. CoPro facilitates time-resolved quantification of protein concentrations in single functional cells, and enables the distributions of concentrations across a cell population to be measured. This could be useful in investigating several cellular processes which are mediated by proteins, especially where changes in protein concentration in a single cell in response to changes in the extracellular chemical environment are subtle and rapid and may be smaller than the variability across a cell population.

## 1. Introduction

A number of different methods already exist for quantifying protein copy number in cells, mainly involving bulk ensemble average biochemical techniques of cell lysates, including tag-affinity quantification and mass spectrometry,[1] but more recently, using fluorescence measurements on living cells. Affinity methods (reviewed here[2]) involve tagging a protein of interest and quantifying it using western blots or ELISA, and have achieved zeptomolar (i.e. 100s molecules/cell) resolution.[3] The entire yeast proteome has been mapped using affinity methods.[4] Mass spectrometry combined with chromatography can generate peptide spectra which can be identified using peptide databases[5] and has been used to map the proteome in the malaria parasite, *Plasmodium falciparum*.[6] Fluorescence detection has been used with flow cytometry to measure the copy numbers of >4,000 GFP tagged proteins in *S. cervisiae*.[7] Comparing the cytometry results with western blots, proteins with >8,000 copies/cell were precisely quantified. However, the sensitivity of detection fell off steeply with lower copy numbers, and at copy numbers equivalent to 2,000-4,000 molecules per cell this sensitivity dropped to ~50%. Fluorescence detection using microscopy methods have enabled more precise quantification - for example a YFP-tagged chromosomally integrated protein library was used by Taniguchi et al. with live cells investigated using fluorescence microscopy combined with automated microfluidics to study the *E. coli* proteome.[8]

In our study here, we have developed a fluorescence microscopy single-molecule tool we call the CoPro (Concentration of Proteins) method to generate robust estimates of protein concentrations inside different subcellular compartments of living cells. To characterize our method we used robust simulations of realistic complex cell shapes and compartments corresponding to typical rod-shaped prokaryotic bacteria and spherical eukaryotic yeast cells. We then applied our approach experimentally using *E. coli* bacterial cells containing fluorescently labelled replisome component, DnaQ, labelled protein to compare against the earlier method of Taniguchi et al, used for their *E. coli* proteome studies. Following this we ultimately applied the method experimentally to a more complex budding yeast cell system for which we could control the protein concentrations in different subcellular compartments by precise manipulation of the extracellular environment. The glucose sensing pathway of budding yeast was an excellent choice in this regard, offering a complex cellular system which has multiple subcellular compartments and with the protein concentrations of key regulators of the signal transduction response known to vary in response to extracellular glucose concentration, from earlier biochemical and standard epifluorescence microscopy studies.

All cells dynamically sense their environment through signal transduction mechanisms, and in the majority of these mechanisms gene regulation is involved to respond to environmental changes (a notable exception being chemotaxis in many bacteria such as

*E. coli* which utilize a protein-only regulatory network in the cell). These gene regulatory mechanisms rely on cascades of protein-protein interactions which transmit signals from sensory elements to responsive elements within each cell. The glucose sensing mechanism in budding yeast, *Saccharomyces cerevisiae*, is a model system for the study of signal transduction which involves gene regulation. The multi-copy inhibitor (a protein called Mig1) of the GAL gene expression protein is an essential transcription factor in this mechanism. Mig1 is a $Cys^2$-$His^2$ zinc finger DNA binding protein[9] which binds several glucose-repressed promoters.[10–13] In the presence of elevated concentration levels of extracellular glucose it is poorly phosphorylated and predominantly located in the nucleus[14,15] where it recruits a repression complex to the DNA.[16] If extracellular glucose concentrations levels are depleted, Mig1 is phosphorylated by the sucrose non-fermenting protein (Snf1)[17–19], resulting in a redistribution of mean localization of Mig1 into the cytoplasm.[14,20,21] Thus, Mig1 concentration levels in the cell nucleus and cytoplasm serve as a readout of glucose signal transduction in budding yeast.

Standard epifluorescence microscopy has been used previously to quantify the ratio of mean fluorescence pixel intensity of GFP tagged Mig1 in the cytoplasm with the nucleus in live budding yeast cells.[15] Bendrioua at al. used microfluidics assays to observe changes in this ratio of mean intensity values in real-time in response to glucose changes. They observed a rapid (<1 min) response of cells to extracellular glucose concentration changes in a history-dependent manner dependent on cells' previous exposure to glucose, indicative of cellular sensory adaption. However, using these methods they were not able to quantify the absolute number of Mig1 molecules in the nucleus or the cytoplasm but relied instead only on the fluorescence intensity as an uncalibrated readout of Mig1 concentration. Here, we have used our home-built single-molecule narrowfield microscope, in combination with automated super-resolution localization microscopy, image segmentation and convolution analysis, to directly quantify the concentration of Mig1 in the nucleus and cytoplasm of live yeast cells at high and low glucose concentrations.

In narrowfield microscopy, the area of a collimated epifluorescence laser excitation field in the focal plane is reduced compared to conventional epifluorescence by an order of magnitude to have a typical width of 5-10 μm which can be concentrated to illuminate just a single cell, similar to Slimfield which has a comparable sized beam waist in the focal plane but is generated using a focused laser beam in the sample.[22] Both narrowfield and Slimfield illumination generate high local excitation intensity fields of typically a few kW $cm^{-2}$, which enable single fluorescent protein detection above camera readout noise at millisecond sampling times, which is required to track the movement of individual molecules diffusing in relatively low viscosity cellular environments, such the cell cytoplasm[23]. We have previously used millisecond imaging to quantify the concentration of bacterial replisome protein components in single *E. coli* bacterial cells.[24] In our study here, we extend these methods to quantify the concentration of fluorescently labelled molecules in more complexly structured yeast cells. These cells have a total volume an order of magnitude greater and have significantly higher level of native autofluorescence. They also contain complex subcellular compartments, potentially with different concentrations in different compartments.

A narrowfield fluorescence image of a cell compartment containing fluorescently-labelled components is comprised of four principal intensity components: 1. a background signal from camera noise; 2. a cellular autofluorescence background from naturally fluorescent molecules which are native to the cell; 3. foreground spots of fluorescence of varying degrees of brightness corresponding to one or more fluorophores which have been introduced in the cell to label specific molecular components – fluorophores which are colocalized in space to within less than the optical resolution limit of 200-300 nm are detected as being part of the same 'fluorescent spot' (for example, a molecular complex of effective diameter of a few nm may be composed of $N$ repeating subunits of the same protein, and if this protein subunit is labelled with a single specific fluorophore, the brightness of the fluorescence spot that we observe in the far-field diffraction-limited regime of our microscope is $N$ times the brightness of a single fluorophore) within the microscope depth of field; 4. a background pool of, typically, rapidly diffusing, fluorescently-labelled molecules that are not detected as distinct spots of fluorescence.

Each of these separate intensity components must be quantified to obtain the total concentration of fluorescently-labelled molecules. The autofluorescence background and camera noise were characterized by narrowfield microscopy images of wild-type parental cells (i.e. cells containing no fluorescent protein labelling). The mean intensity of the autofluorescence was subtracted from the fluorescence signal to obtain the fluorescent protein foreground signal. The number of molecules in distinct 'spots' or 'foci' of fluorescence, which have a mean effective diameter of a few hundred nm consistent with the measured point spread function (PSF) width of our microscope, was measured using our bespoke super-resolution localization microscope[25,26] which could objectively track automatically detected candidate fluorescent spots over time using robust probabilistic criteria. The brightness of these spots can be compared against a separate calibration obtained for the brightness due to a single fluorophore (for example, one can measure step-like intensity changes due to photobleaching of single fluorescent protein molecules obtained in separate *in vitro* experiments in which single fluorescent protein molecules have been immobilized via specific chemical conjugation to a microscope coverslip surface).

## 2. Materials and Methods

### 2.1 Strains and growth conditions

We used a DnaQ-GFP *E. coli* strain, adapted from DnaQ-YPet[24] by PCR amplification of eGFP with A206K mutation[27] for monomerization, and *kan I* using primers with 50 nt overhangs homologous to the insertion region. *E.coli* strains were grown in 56 salts minimal media supplemented with 4% glucose and grown overnight at 37˚C. We used the MATa MIG1-GFP-HIS3 NRD1-mCherry- hphNT1METLYS *S. cerivisiae* strains in the BY4741 background.[15] Yeast strains were grown in YNB minimal media supplemented with 4% glucose overnight at 30 ˚C.

## 2.2 Purification of fluorescent proteins

His-tagged mCherry and GFP genes were amplified by PCR and cloned into PET vectors. These were transformed into BL21 PLysS *E. coli* and grown to saturation overnight in 2l 2YT media supplemented with 20 mg/mL kanamycin. Overnight cultures were induced with 100 µg/ml IPTG and allowed to express protein for 5 hours at 20˚C. Induced culture was pelleted by centrifugation, resuspended in lysis buffer (10% glycerol, 50 mM Tris-HCl, 150 mM NaCl, 10 mM imidazole, 1 mM DTT, pH 8) and lysed by sonication. Lysate was cleaned by centrifugation and the supernatant applied to pre-packed NTA columns (His Gravitrap, GE Healthcare) equilibrated with lysis buffer at 4˚C. The column was washed with lysis buffer before protein was eluted with lysis buffer supplemented with 500mM Imidazole. 1 mL fractions were collected and run on a denaturing gel. Fractions containing the protein were pooled and dialysed overnight in storage buffer (50% glycerol, 50 mM Tris-HCl, 150 mM NaCl, 10 mM imidazole, pH 8). The protein was then aliquoted and flash frozen in liquid nitrogen to be stored at -80 ˚C.

## 2.3 Narrowfield microscope

A bespoke inverted fluorescence microscope was constructed using a Zeiss microscope body with a 100x TIRF 1.49 NA Olympus oil immersion objective lens and a *xyz* nano positioning stage (Nanodrive, Mad City Labs). Fluorescence excitation used 50mW Obis 488nm and 561nm lasers. A dual pass GFP/mCherry dichroic with 25nm transmission windows centred on 525nm and 625nm was used underneath the objective lens turret. The beam was expanded 0.5x and 1x for imaging bacteria and yeast cells respectively with a series of lenses on selectable flipper mounts, to generate an excitation field of intensity ~6 Wcm$^{-2}$. Beam intensity profiles were measured directly by raster scanning in the focal plane while imaging a sample of fluorescent beads. A high speed camera (iXon DV860-BI, Andor Technology, UK) was used to image at typically 5ms/frame (this rapid sampling speed was required to image diffusing molecules in the cell) with the magnification set at ~80 nm per pixel. The camera CCD was split between a GFP and mCherry channel using a bespoke colour splitter consisting of a dichroic centred at pass wavelength 560 nm and emission filters with 25 nm bandwidths centred at 525 nm and 594 nm. The microscope was controlled using our in-house bespoke LabVIEW (National Instruments) software.

## 2.4 *In vitro* microscopy

*In vitro* experiments were performed in a simple tunnel slide flow-chamber constructed from strips of double-sided tape creating a channel on a standard glass microscope slide and covered with a plasma-cleaned BK7 glass coverslip, creating a chamber 5-10 µl in volume, using a protocol adapted from earlier studies.[28,29] In brief, the PSF was measured using 20 nm diameter fluorescein beads (Invitrogen) diluted by a factor of 1,000 in PBS and one volume injected into the flow chamber. Beads were left to sediment onto the coverslip for 10 min with the chamber inverted before excess beads were washed out with 10 flow-chamber volumes of PBS. *In vitro* fluorescent proteins, GFP and mCherry were imaged by flowing one flow-chamber volume of 1 µg/ml anti-GFP or anti-DsRed antibodies respectively. After 5 min incubation at RT in the inverted flow-chamber excess antibody was washed away with 10 flow-cell volumes of PBS. One flow-cell volume of 1 µg/ml fluorescent protein was then injected, incubated for 5 min and washed. To focus on the coverslip surface in brightfield, 1,000-fold dilution of 300 nm diameter polystyrene beads (Invitrogen) was added to the slide, incubated and washed.

**2.5 *In vivo* microscopy**

Budding yeast and *E. coli* cells were imaged on agarose pads.[24] In brief, gene frames (Life Technologies) were stuck to a glass microscope slide to form a well and 500 µl YNB or 56 salts, for yeast or *E. coli*, plus 1% agarose pipetted into the well. The pad was left to dry at room temperature before 5 µl overnight yeast or *E. coli* culture was pipetted in 6-10 droplets onto the pad. This was covered with a plasma-cleaned glass coverslip and imaged immediately. The overnight yeast culture was used as high (4%) glucose concentration and 4% glucose included in agarose pad. For low glucose concentrations, the overnight culture was spun down, washed and resuspended in YNB with no glucose. Imaging consisted of finding a cell in brightfield mode, recording a stack of 10 brightfield images before taking stacks of 100-1,000 frames of fluorescent images for each fluorescent channel separately.

**2.6 Image analysis**

The diffusive pool fluorescence (due to particles which diffused too rapidly, and/or which were too dim, and/or which were too close together to be detected as a distinct fluorescent spot) was modelled as a 3D convolution integral of the normalized PSF, *P*, of our imaging system, over the whole cell. Figure 1 illustrates the model diagrammatically in the case of a simple spherical cell. Each camera detector pixel of physical area Δ*A* has an equivalent area *dA* in the conjugate image plane of the sample mapped in the focal plane of the microscope, such that *dA=ΔA/M* where *M* is the total magnification between the camera and the sample. The measured intensity, *I'*, in a conjugate pixel area *dA* is the sum of the foreground fluorophore intensity *I* plus autofluorescence ($I_a$) plus detector noise ($I_d$). *I* is the is the sum of the contributions from all of the non-autofluorescence fluorophores in the whole of the cell, such that:

$$I(x_0, y_0, z_0)dA = \sum_{i=1}^{AllCellVoxels} E\rho I_s P(x_i - x_0, y_i - y_0, z_i - z_0,) \qquad (1)$$

Here $I_s$ is the characteristic integrated intensity of a single fluorophore and $\rho$ is the fluorophore density in units of molecules per voxel (i.e. a pixel volume unit). $E$ is a function representing the change in the laser profile excitation intensity over the cell. In a uniform excitation field $E=1$. For narrowfield microscopy the excitation intensity is uniform in $z$ but has a 2D Gaussian profile in the lateral $xy$ plane parallel to the microscope focal plane. In a non-saturating regime for photon emission flux of a given fluorophore, the brightness of that fluorophore, assuming simple single-photon excitation, is proportional to the local excitation intensity[24], thus:

$$E(x, y, z) = \exp\left[-\left(\frac{x^2 + y^2}{2\sigma_{xy}^2}\right)\right] \qquad (2)$$

Here $\sigma_{xy}$ is the Gaussian width of the excitation field in the focal plane, which we configure to be 3 µm and 6 µm in our microscope for the *E.coli* and budding yeast configurations respectively. In Slimfield there is a $z$ dependence also with Gaussian sigma width which is ~2.5 that of the $\sigma_{xy}$ value, but in narrowfield it is independent of $z$, which made narrowfield a more ideal choice of illumination here for larger cells such as yeast. Thus:

$$I(x_0, y_0, z_0)dA = \sum_{i=1}^{AllCellVoxels} \rho I_s \exp\left[-\left(\frac{x^2 + y^2}{2\sigma_{xy}}\right)\right] P(x_i - x_0, y_i - y_0, z_i - z_0,) \qquad (3)$$

Defining $C(x_0,y_0,z_0)$ as the numerical convolution integral (also containing the Gaussian excitation field) over the specific cell being imaged, and assuming the fluorophore density averaged over time is uniform in space in any given subcellular cell compartment, we can calculate the fluorophore density from each pixel in the image as:

$$\rho = \frac{I(x_0, y_0, z_0)dA}{C(x_0, y_0, z_0)I_s} = \frac{I'(x_0, y_0, z_0) - (I_a + I_d)}{C(x_0, y_0, z_0)I_s} \qquad (4)$$

Here, the mean value of the total background noise ($I_a + I_d$) can be calculated from images of cells which do not contain any foreign fluorophores. To calculate the mean value of the fluorophore density in the diffusive pool of fluorophores in a given cell compartment we can average over all $\rho$ estimates corresponding to all pixels inside that compartment whose boundaries have been determined by automated image segmentation and whose pixels are not associated with distinctly detected spots of fluorescence due to tracked assemblages of fluorophores (e.g. a molecular complex). This generates a robust estimate for fluorophore density in the diffusive pool which is unique not just to a specific single cell, but also provides a rough estimate to a specific single compartment within that cell. This rough estimate is useful in providing a simple preliminary and computationally non-intensive quantitation because it assumes no prior knowledge of subcellular structures which the cell.

In a more general case of multiple cellular compartments of different sizes, locations and shapes we can extend this model, assuming that the time-average pool concentration within each given intracellular compartment (for example, the nucleus, the cytosol compartment, and many other types of cell organelle) can be characterized by a mean value subject to small fluctuations across the extent of the compartment. However we do not assume that the mean values within different cellular compartments are necessarily equal:

Thus, in the case of two such cellular compartments containing the cytoplasm and the nucleus we have:

$$I(x_0, y_0, z_0)dA = I_s \exp\left[-\left(\frac{x^2 + y^2}{2\sigma_{xy}}\right)\right]\left(\begin{array}{c}\sum_{i=1}^{CytoplasmVoxels}\rho_c P(x_i - x_0, y_i - y_0, z_i - z_0,) \\ + \sum_{i=1}^{NucleusVoxels}\rho_n P(x_i - x_0, y_i - y_0, z_i - z_0,)\end{array}\right) \quad (5)$$

Here, $\rho_c$ and $\rho_n$ refer to the mean cytoplasmic and nuclear concentrations respectively. By analysing all pixel data from both compartments we can then use least-squares regression analysis in Matlab to estimate mean values for $\rho_c$ and $\rho_n$. This can be generalised to any number of intracellular compartments:

$$I(x_0, y_0, z_0)dA = I_s \exp\left[-\left(\frac{x^2 + y^2}{2\sigma_{xy}}\right)\right]\sum_{j=1}^{Numberof\ Compartments}\sum_{i=1}^{Compartment\ Voxels}\rho_j P(x_i - x_0, y_i - y_0, z_i - z_0,) \quad (6)$$

Where $\rho_j$ is mean concentration of the *jth* compartment. This has an advantage of taking into account the full contribution of all compartments to each observed pixel intensity in the image. These values can then be used to generate a modified convolution integral *I'* in Equation 4, thus allowing the pixel-by-pixel variation of $\rho$ to be estimated. An important feature of this general model is that each separate compartment does not necessarily have to be modelled by an ideal geometrical shape (such as a sphere, for example) but can be any enclosed 3D volume provided its boundaries are well-defined to allow numerical integration in Equation 6.

**2.7 Analytical PSF**

The 3D PSF can be approximated as a series of 2D Fourier transforms of the pupil function at different *z* slices[30]:

$$P(x, y, z) = \iint_{k_x k_y} F(k_x, k_y, z)\exp(j(k_x x + k_y y))dk_x dk_y \quad (7)$$

Where $k_x$ and $k_y$ are the 2D Fourier coordinates and $F(k_x, k_y, z)$ is the pupil function which describes the field distribution in the pupil of the objective lens, When $\sin\theta_i = NA/n_i$, $F$ is given by:[31]

$$F(k_x, k_y, z) = A(\theta_i) \exp(jk_0 \phi(\theta_i, \theta_s, z)) \tag{8}$$

Otherwise, $F = 0$. The angles $\vartheta_{i,s}$ can be defined in Fourier coordinates (see Figure 2):

$$\sin\theta_{i,s} = \frac{k_x^2 + k_y^2}{k_{i,s}} \tag{9}$$

Here $k_{i,s}$ is the wavenumber, $k_{i,s} = 2\pi n_{i,s}/\lambda$, of the emitted light (we approximate the wavelength to the peak emission wavelength, $\lambda$=515 nm for GFP and 610 nm for mCherry) through the immersion medium (refractive index $n_i$ =1.515), specimen (refractive index $n_s$ =1.33) or $k_o$ is the wavenumber through a vacuum and NA is the numerical aperture of the objective lens (in our study here, NA=1.49). $A(\vartheta_i)$ is the apodization function, for an emitting point source such as a fluorophore:[32]

$$A(\theta_i) = (\cos\theta_i)^{-\frac{1}{2}} \tag{10}$$

The phase function $\phi(\vartheta_i, \vartheta_s, z)$ describes optical path difference of a wavefront exiting the pupil compared to a reference wavefront and is the sum of a defocus term $\phi_d$ and an aberration term $\phi_a$. The defocus term is approximated by:[30]

$$\phi_d \approx n_i z (1 - \cos\theta_i) \tag{11}$$

And the aberration term, caused by differences in the immersion and sample media can be derived from geometric optics as:

$$\phi_a = d(n_s \cos\theta_s - n_i \cos\theta_i) \tag{12}$$

Where $d$ is the distance from the front surface of the objective lens to the focal plane (~400 μm in our case).

**3. Results and Discussion**

**3.1 Estimating the PSF**

To perform the convolution analysis, the PSF of the microscope was first experimentally estimated. Here, we measured the PSF of the our microscope using narrowfield illumination by acquiring experimental fluorescence images of 20 nm diameter green fluorescent beads (Invitrogen) immobilized to the glass coverslip surface of our microscope flow-chamber.[33] The beads were imaged at 100 nm intervals in *z* which was controlled by an automated piezo nanostage (Nanodrive, Mad City Labs) controlled by our own bespoke software (LabVIEW, National Instruments) over a *z*-range ±1 µm above/below the focal plane. Time-series images of six different beads using the same imaging conditions as for live cell microscopy were averaged pixel-by-pixel to form the experimental PSF (Figure 3), which was then background-corrected by subtracting the mean local background and then normalized by dividing by the total summed pixel intensities of the detected foreground spot image. At absolute values of *z* beyond ~1 µm from the focal plane, the weak defocused intensity fluorescence signal from single beads was difficult to resolve above camera noise. However, by modelling the experimental PSF data obtained over the ±1 µm *z*-range using an analytical PSF formulation we could then extend the range of the PSF determination to cover the full extent of single yeast cells whose diameter is typically ~5 µm, but can be as high as ~10 µm at certain stages in the cell cycle for budding yeast cells

For this analytical approximation we used a Matlab implementation[34] of the Stokseth method[30] (Figure 3), outlined in the Material and Methods section 2.7. This models the PSF as the Fourier transform of the pupil function which describes the field distribution in the pupil of the objective lens for given wavelengths of light.

The analytical function and experimental data were found to be in good agreement to the experimental PSF, with an associated mean *chi* squared value of ~65 when measured on a pixel-by-pixel basis across each PSF image equivalent to a probability confidence interval $P<0.001$. Figure 3 shows a qualitative comparison in which we have added realistic noise to the analytical PSF.

**3.2 Fluorescent proteins *in vitro***

To test the single-molecule detection capabilities of the narrowfield microscope and to measure the characteristic intensity of single fluorophores, we imaged purified GFP (Clontech eGFP, with the addition an A206K mutation to inhibit GFP dimerization[27]) and mCherry (Clontech) immobilized to the coverslip surface via antibody conjugation. Representative images of GFP and mCherry are shown in Figure 4a and b left panel. Rapid photobleaching occurred within only a few image frames so the first frame is shown. Bright spots were tracked over time using millisecond sampled images, analysed using our custom Matlab software.[25,26] The software objectively identifies candidate bright spots by a combination of pixel intensity thresholding and image transformation. The threshold is set using the pixel intensity histogram as the full width half maximum of the peak in the histogram which corresponds to background pixels. A series of erosion and dilation is applied to the thresholded image to remove individual bright pixels due to noise. A final erosion step then leaves a single pixel at each candidate spot co-ordinate. The intensity centroid and characteristic intensity, defined as the sum of the pixel intensities inside a 5 pixel radius region of interest around the spot minus the local background[35] and corrected for non-uniformity in the excitation field are determined by iterative Gaussian masking[36] which resulted in a mean localization precision of ~40 nm and ~55 nm for GFP and mCherry molecules respectively. Localisation precision was measured as the standard deviation of the spot centroid over time. Spots are accepted as real if their signal-to-noise ratio is above a threshold which was pre-determined from simulated data, using a realistic noise distribution. This threshold initially was set generously, equivalent to a level of 35% false positive detection probability per image frame, however additional tracking criteria for subsequent image frames (spots are only ultimately accepted if they constitute a track or trajectory, meaning: i, they last for at least 3 consecutive image frames; ii, the intensity centroid displacement between spots in consecutive images is 5 pixels (one spot region of interest) or less; iii, the width and integrated intensity of a spot in an image frame in a given track is within 50% of that measured in the previous image frame). These additional tracking acceptance criteria reduced the likelihood of false positive detection to <1%.

Spots are linked into trajectories based on their proximity to neighbouring spots in subsequent image frames, their integrated intensity and their estimated size based on a 2D unconstrained Gaussian fit to their experimental PSF intensity profile. Figure 4 middle panel shows the intensity as a function of time of GFP and mCherry spots which were 'overtracked' (i.e. where the integrated intensity continued to be measured at the intensity centroid to visually indicate the level of local background intensity in the absence of the spot), after a spot had photobleached to zero mean integrated intensity. These traces are overlaid using a bespoke Chung-Kennedy edge-preserving filter[37,38] and show a step-like drop to mean zero intensity, indicative of a molecular signature for single fluorescent proteins.  Kernel density estimations (KDEs) were used to obtain the distribution of all integrated spot intensity measurements of GFP and mCherry images.[39] This involves a 1D convolution of the spot integrated intensity data with a Gaussian kernel of unitary integrated area (i.e. equivalent to a total of just one data point) and an optimised bandwidth of ~600 counts which was determined objectively from the software, and are shown on the right panels of  Figure 4. This approach results in significant objectification of the displayed distribution in comparison to standard histogram methods.

The distributions peak at ~4,400 counts and ~3,400 counts on our EMCCD detector for GFP and mCherry respectively which we use as best estimates for the characteristic brightness values. We use these estimates as opposed to mean values from the distributions which are biased towards marginally higher values than the peak values due to a tail on the distribution comprised in part from a minority of detected spots which include two or more individual fluorescent protein molecules whose separation of the coverslip surface is less than the optical resolution limit. Step-wise photobleaching data measured from live budding yeast cells indicated similar levels of brightness to within ~10% consistent with earlier in vivo single-molecule studies of *E. coli* bacteria using fluorescent proteins.[24,28,29]

### 3.3 Localisation precision

Since localisation microscopy is required in the estimation of diffusive pools pixels in a given compartment we sought to characterize the localisation precision of our narrowfield illumination single-molecule microscope over a range of different fluorescent spot contrast values. As a model for bright fluorescent spots we captured time series of surface-immobilised 20 nm green fluorescent beads (Invitrogen) which had been exposed to varying pre-bleach laser exposures. By varying the bleach times we were able to generate images of spots over a range of effective spot signal-to-noise ratios. Spots were tracked as before and their localisation precision determined as the standard deviation in intensity centroid position over time. Localisation precision against the signal-to-noise ratio is plotted in Figure 5 in blue. The data was fitted using the Thompson equation[36] for localisation precision of 2D data, given by:

$$\left\langle (\Delta x)^2 \right\rangle = \frac{s^2 + a^2/12}{N} + \frac{8\pi s^4 b^2}{a^2 N^2} \quad (11)$$

Here, $s$ is the sigma width of the PSF (170 nm for green fluorescent beads), $a$ is conjugate equivalent pixel size in the sample focal plane (80 nm), $N$ is number of photons collected belonging to spot and $b$ is the number of background photons collected. The intensity, $I$, collected from a given detected spot of fluorescence is equal to the number of photons emitted from the spot multiplied by a constant $G$ which incorporates camera gain and total photon collection efficiency between the sample and the camera detector. The signal-to-noise ratio was defined as the integrated spot intensity divided by the standard deviation of the local background of the spot (calculated from pixels which are within a square 17x17 pixel array centred on the intensity centroid of the spot but excluding pixels that are contained within the central 5 pixel radius circular region of interest which comprises of the spot integration area), equivalent to ~8 counts for the fluorescent bead data, constant for each set of imaging conditions) multiplied by the area of the spot (80 pixels). The Thomson equation was fitted to the spot intensity data to generate the localisation precision as a function of signal-to-noise ratio for fluorescent nanobead data, shown in black in Figure 5 with 90% confidence intervals as dotted lines, using an optimised collection constant of $G$=0.1 and background photon count, $b$=5 photons. The localisation precision of GFP and mCherry is also shown in Figure 5 in green and red respectively (note, since mCherry has a larger PSF width than the green fluorescent beads of the GFP there is a marginal deviation from the fit extrapolated from the green bead data).

## 3.4 Concentration measurements in *E. coli*

As proof-of-principle we tested the CoPro method on the single cytoplasmic cellular compartment of model *E. coli* bacteria using a cell strain consisting of the replisome protein DnaQ fused to GFP using chromosomal integration. This was essentially identical to an earlier cell strain developed, but which instead used the yellow YPet fluorescent protein as the fluorophore tag. This cell strain's DnaQ protein copy number per cell had previously been estimated using both quantitative western blots and a different convolution method using Slimfield illumination which used different experimentally derived PSF estimates.[24] We took narrowfield fluorescence images of the DnaQ-GFP tagged cell strain and quantified the cellular concentration of DnaQ.

Figure 6 shows images of a representative DnaQ-GFP cell and quantification of DnaQ concentration. A brightfield image of a cell is shown in grey in Figure 6a, with the segmented outline of the cell which was obtained from the raw cell fluorescence image using an automatically varying threshold based on the pixel intensity distribution shown overlaid in orange. This raw cell boundary image segmentation, in the case of *E. coli* cells, could be modelled as a 'sausage' shape (shown overlaid in white on the raw fluorescence image shown in green in Figure 6b). The segmentation threshold was configured to correspond to low intensity autofluorescence which was observed to be delocalized in the cytoplasm, and so was not restricted to the spatial distribution of GFP-tagged material in the cytoplasm compartment. The centroid, orientation and major and minor axis lengths of the cell area were used to define the unique fitted 'sausage' function around each cell, which consisted of a rectangle capped by a half-circle at either end,[40] which was an accurate 2D projection of the 3D *E. coli* cell shape of a cylinder capped by two hemispheres.[41] Bright detected distinct spots in the fluorescence image were found using the same methods as for the *in vitro* data, and are shown overlaid in Figure 6b as white circles.

The fluorescence pixels associated just with the diffusive pool are shown in Figure 6c and were determined by taking only pixels within the segmented cell area and removing pixels in a 5 pixel radius around the bright spot centroids (i.e. corresponding to the same region of interest area used for the integrated spot intensity determination). The number of DnaQ molecules in bright spots was determined by dividing the total intensity in the 5 pixel radius around the spot by the intensity of a single *in vitro* GFP and was typically <10 molecules. The mean autofluorescence background and camera detector noise background intensity values were also subtracted from all pool pixel intensity values. The convolution integral, *C*, shown as a heat map in Figure 6d, was obtained by integrating the analytical PSF over the sausage function which defined the cell boundaries. A map of the diffusive pool concentration was obtained by dividing the pool pixel intensity values by the corresponding convolution integral pixel values, on a pixel-by-pixel basis, and by the characteristic peak integrated intensity for a single molecule GFP obtained from the *in vitro* surface-immobilized assay. The pixel map for this calculation is shown in Figure 6e in units of molecules/voxel.

The KDE of cell DnaQ protein concentration sampled from a population of only seven cells is shown in Figure 7, showing that a probability distribution for a cell population can be generated with this method using relatively low numbers of cells. The overall effective DnaQ concentration in the entire compartment was defined as the mean DnaQ protein concentration in the diffusive pool plus the total amount of DnaQ detected in any distinct fluorescent spots divided by the effective volume of the cellular compartment, and was used to determine the total copy number of DnaQ molecules in each individual cell by multiplying the estimate of the overall effective DnaQ concentration by the calculated volume given by the cell's unique sausage function. The mean copy number of DnaQ ± standard deviation was calculated to be 350±120 molecules which agrees well with the copy number of DnaQ-YPet, measured previously using the YPet fluorescent protein and a different convolution method, of 270±160 molecules.[24] The distribution of DnaQ concentration per cell is broad, reflecting cell-to-cell variation – this information is lost when quantifying protein concentrations in cells using traditional bulk ensemble biochemical assays. These variations may be caused by each cell being at a different phase in the cell cycle but may also reflect that each cell is an individual and, to fully understand cellular behaviour, this individuality must be characterized.

**3.5 Concentration measurements of simulated structured cells**

To extend our method to more complexly structured cells, we first tested it on realistic simulated images from different cell types containing multiple cellular compartments. Figure 8a shows two simulated budding yeast images. The left panel indicates a simulated nuclear concentration of Mig1 which is double the cytoplasmic concentration. Figure 8a right panel is a simulated image in which the nuclear concentration is zero with all the simulated Mig1 molecules localized uniformly to the cytoplasm. The cell periphery is labelled in orange and the nuclear periphery in cyan. The cell and the nucleus were modelled as convolution integrals of spheres with radii of 32 and 15 pixels respectively containing uniform fluorophore concentrations. Realistic noise was added in the lower panel by adding normally distributed random pixels with a standard deviation ~20% peak cytoplasm intensity. Images were simulated over a range of nuclear concentrations from zero to five times a constant cytoplasmic value.

Each pixel intensity value is the sum of the convolution integral of each separate compartment in the cell (Equation 5) so the concentration in each compartment was determined by solving a set of simultaneous equations for each pixel value using our bespoke Matlab-coded software. Equations were solved by linear least squares regression analysis with the only constraint that concentration cannot be negative. Figure 8b shows the measured cytoplasmic concentration as determined by our CoPro algorithm on the simulated data against the simulated nuclear concentration, with errorbars given by the standard deviation over five repeated simulations. Figure 8c shows the same for the nucleus with an insert to show an expanded section of the data and the fit. The linear fits of Figure 8b and 8c are associated with chi squared values equivalent to a probability confidence interval $P<0.0001$. These simulations thus imply the method is very robust in yeast cells over a very broad range of relative compartment concentrations.

To demonstrate the generality of our method we also simulated a different cell type with multiple different sized components and different locations, shown without and with noise in Figure 9 left and middle. The cell is rod-shaped, like an *E. coli* bacterium, and contains three spherical compartments of different volumes and internal protein concentrations. Although not modelled on any particular system, this cell resembles, for example, fluorescently labelled carboxysomes in cyanobacteria. The mean concentration in each compartment across five repeated simulations was again calculated by solving the simultaneous equations from four separate convolution integrals at each pixel and is plotted against the simulated concentration in Figure 9. The measured concentrations all agree very well with the simulated concentrations, demonstrating this method is also robust in this example of non-spherical cells containing multiple different cellular compartments.

### 3.6 Concentration measurements in budding yeast

We then applied these methods to quantify experimentally the concentration of Mig1 protein molecules in budding yeast cells at high and low glucose conditions, and also to estimate the concentration of the RNA polymerase protein Nrd1 as a control, since Nrd1 is not directly involved in the glucose sensing pathway. Figure 10 shows images of a representative dual-label Mig1-GFP:Nrd1-mCherry cell with the corresponding quantification of Mig1 concentration. Figure 10a shows a brightfield non-fluorescence image of the cell with the nuclear membrane and cell membrane boundaries overlaid in cyan and orange respectively. The algorithm used for image segmentation was similar to that employed for *E. coli* above, based on the fluorescence image of Mig1 (shown in Figure 10b) for the cell boundary and Nrd1 (shown in Figure 10c) for the nucleus, as before configuring threshold levels to be sensitive to delocalised cellular autofluorescence.

Bright spots were tracked as before, and shown overlaid in white on Figure 10b. Due to much higher levels of cellular GFP fluorescence and autofluorescence, spots are harder to display in the image, so a zoomed-in cut-out on the figure shows a typical fluorescent spot with the intensity display levels adjusted appropriately. The nuclear and cytoplasmic concentrations were determined separately. The distinctly detected spots of fluorescence were removed from the fluorescence images and background correction applied. The concentration in the nucleus and cytoplasm were determined as for the simulated images, modelling the cell and nucleus as spheres with radii determined from the segmentation. Figure 10d shows a spatial map of the pool concentration fluctuation obtained by dividing the background corrected fluorescence image by the sum of the convolution integrals in the cytoplasm and nucleus multiplied by their concentration value and the characteristic GFP intensity. Each pixel in the fluctuation map represents the percentage difference from the compartment mean.

The KDE distribution of Mig1 concentration in the nucleus and cytoplasm at high and low glucose is shown in Figure 11 left and middle. In high levels of extracellular glucose concentration, the concentration of Mig1 is much higher (roughly by a factor of 4) in the nucleus than in the cytoplasm. In low extracellular glucose concentrations the concentration of Mig1 is more similar in the nucleus and cytoplasm but still elevated in the nucleus by ~30%. As a control, we measured the concentration of the Nrd1 protein at high and low extracellular glucose concentrations. Nrd1 was found to be almost exclusively localized in the nucleus, with the peak in the Nrd1 concentration distribution indicating a copy number of ~2,000 molecules, and the distributions for Nrd1 nuclear concentration are shown in Figure 11 right and were shown by Student t-tests to be independent of glucose concentration.

The mean copy number of Mig1 molecules in the nucleus and cytoplasm is shown in Table 1, indicating mean and standard deviation values. Using bulk ensemble average affinity methods, the total copy number of Mig1 in the whole cell was estimated to be ~830 molecules/cell[4] which agrees well with our results – the authors' conservative assessment of error on this copy number estimate was ~100%, which illustrates one of the key advantages of our single-molecule method. The Mig1 concentration is higher in the nucleus than the cytoplasm at low glucose concentrations (~30%), although by much less than at high glucose concentrations (~400%). This suggests that some Mig1 molecules interact in the nucleus even at low levels of extracellular glucose concentration.

|  | Mean number of Mig1 molecules per cell | Standard Deviation (molecules per cell) |
| --- | --- | --- |
| **High Glucose Cytoplasm** | 542 | 200 |
| **High Glucose Nucleus** | 249 | 88 |
| **Low Glucose Cytoplasm** | 1070 | 400 |
| **Low Glucose Nucleus** | 141 | 57 |

Table 1: Copy number of Mig1 molecules, rounded to nearest molecule, in the nucleus and cytoplasm

### 4. Conclusions

The CoPro method for determining the protein concentration in live cells was first tested using the DnaQ protein concentration in *E.coli* as a previously studied system, and found to produce similar results to within experimental error. In applying CoPro to budding yeast cells we were able to quantify changes in protein concentration in cellular compartments in response to controlled environmental changes; here, by changing the extracellular glucose concentration and then using CoPro to monitor the protein concentration of the protein Mig1 which performs a biological role as a response regulator in the glucose sensing pathway in yeast. By measuring the concentration of Mig1 in both the nuclear and cytoplasmic compartments in each budding yeast cell, cell-by-cell, we were able to obtain distributions across cell populations, enabling observation of subtle concentration shifts. These results show promise for the investigation of future biological systems which may exhibit relatively small changes in concentration of protein in a particular cellular compartment for a given cell, which may be smaller than the variability across the whole cell population and thus hidden were traditional ensemble average approaches to be used in the assessment of protein copy numbers in cells. These distributions of protein concentration also render probabilistic information for the number of a specific protein type in a cell, and how they are distributed spatially between different regions of the cell, which is invaluable information that can be correlated back to stochastic models for gene expression activity.

Importantly, the CoPro method is entirely general in regards to the shape and size of different cellular compartments. Although the case of budding yeast cells involves ostensibly spherical nuclei and cells, the algorithm only requires that the 3D volume of each cellular compartment is well-defined. In the case of hypothetical, asymmetrical cellular compartments which are difficult to model as ideal geometrical shapes, for example, the appropriate numerical integrations can still be performed provided sensible physical estimates of the compartment shapes and relative orientations can be made. These, for example, could be enabled using a separate imaging technique in separate experiments, such as transmission electron microscopy of thin sample sections. Or, utilizing confocal microscopy to obtained z-sections though the cell. The time resolution of typical confocal microscopy is not high enough to follow molecular motions inside living cells, however it can be used potentially to determine the complex 3D shapes of appropriate cellular compartments in conjunction with CoPro which then has the rapid time resolution to map out spatial fluctuations in protein concentration in real time on each separate image frame obtained from millisecond single-molecule microscopy.

Although fluorescence microscopy has been used for quantitative bacterial proteomics previously, both from our own earlier work[24] and those of others,[8] here we have demonstrated a new method which can be applied to larger, more complexly structured eukaryotic cells to quantify protein concentration distributions at a subcellular level. Budding yeast cells have a well-defined approximately spherical structure, allowing for simple convolution integrals but using numerical integration and 3D microscopy techniques such as light sheet microscopy, more complex eukaryotic cells could in principle be investigated. Although the cellular locations used here were the nucleus and cytoplasm, these methods could similarly be applied to other cellular organelles, including mitochondria or vacuoles. Our method also utilises high-speed narrowfield microscopy and can thus be readily combined with single particle tracking to fully quantify protein dynamics in living cells.

In our study here we have used a bespoke narrowfield laser illumination technique which enables millisecond fluorescence microscopy at a single-molecule precise detection sensitivity level. This rapid sampling rate is comparable to the mobility time scale of single proteins in low viscosity cellular environments. For example, a typical fluorescently-labelled protein in the cell cytoplasm has an apparent diffusion coefficient equivalent to a few $\mu m^2/s$, implying that it will diffuse its own point spread function width of the associated fluorescent 'spot' image after just a few milliseconds of observation. This therefore sets a benchmark for the maximum permitted camera exposure time for a single image frame, as above this level such a fluorescent spot appears significantly blurred in a typical image frame and so will fail to be detected as a distinct spot.

Diffusion, however, is a stochastic process; therefore, some single protein molecules may still diffuse greater distances on some given individual image frames and so will fail to be detected, depending also on whether their nearest-neighbour mean separation is less than the optical resolution limit of ca. 200-300 nm. Molecular complexes containing more protein molecule subunits have a higher molecular weight and are likely to have a larger effective Stokes radius and thus lower diffusion coefficient, in addition to their associated fluorescent spots being brighter. Therefore, utilizing CoPro with rapid millisecond imaging enables *separate* experimental quantification between the effective compartment concentrations of proteins present in distinct molecular complexes and those present in lower stoichiometry states that diffuse more rapidly. However, the algorithms of the CoPro method will still work with less rapid imaging rates than those we use here, if these are not technically feasible on a given fluorescence microscope setup; the resultant analysis output will simply indicate fewer, or potentially no, distinctly detected fluorescent spots, but rather output an increased proportion of protein in the diffusive pool in a given compartment. This reduces some aspects of biological insight in regards to lacking the capability to infer the presence or not of distinct molecular complexes in the protein population, but still results in robust quantitative estimates for the total effective numbers of protein monomer units within a cell compartment. A reduced imaging rate also reduces the capacity for *time-resolved* measurements of protein concentration in living cells; instead, slow imaging generates *steady-state* information, but which still has utility in being quantitative on a cell-by-cell and compartment-by-compartment basis.

Our method here utilizes the *in vitro* estimate for the single-molecule brightness of GFP and mCherry. We measured this as being within ~10% of the equivalent *in vivo* brightness in budding yeast cells, consistent with earlier stoichiometry studies of *E. coli* molecular complexes using step-wise photobleaching analysis of fluorescent proteins.[24,28,29] However, it may be possible that in some specialized cellular compartments there is a significantly different pH to the rest of the cell, a good example of which might be lysosomes. In such a case large differences in pH may be manifest as more significant difference in the brightness of single fluorescent proteins in that compartment compared to the rest of the cell. In this circumstance the CoPro method could still be utilized with the modification of a different equivalent brightness value for separate compartments, which could quantified using similar step-wise photobleaching procedures outlined here but pooling statistics into separate distinct compartments on the basis of automated image segmentation.  Not also that although we use fluorescent proteins as reporter labels the CoPro method can generalise to other fluorescent labels; these may ultimately be selected to have less sensitively to changes in local cellular pH, and indeed may also be brighter than fluorescent proteins and have an improved associated localisation precision due to a greater signal-to-noise ratio as evidenced by Figure 5.

Potential issues of using other fluorophores tags beyond fluorescent proteins however include possible homo-FRET/quenching effects. In our study here we could detect no significant correlation between the stoichiometry of tracked fluorescent spots (of Mig-GFP) and the size of the single-molecule photobleach step of the fluorescent protein label (here of GFP), consistent with the earlier single-molecule fluorescent protein stoichiometry studies alluded to previously. This is indicative of an absence of any measurable homo-FRET or quenching effect. The Förster radius of a fluorescent protein FRET pair is in the range 4-5 nm, whereas the closest two florescent protein molecules can physically get to each other is a comparable distance due to the steric hindrance from their beta barrel structure. This indicates that non-radiative energy transitions due to the interaction of electrons in molecular orbitals, whether due to hetero- or homo-FRET, have a relatively small associated signal – the paucity of published single-molecule FRET studies using fluorescent protein FRET pairs lies in testament to this. However, hypothetical quenching  may of course be measureable from smaller dyes were they to be used, and so these effects may need to be characterized in order to minimize associated errors on stoichiometry and protein concentration measurements.

The methods outlined here illustrate not only how light microscopy has evolved from a qualitative observational tool into a highly quantitative instrument,[23] but also how bespoke tools from physics can be developed to characterize properties of the living component of soft-matter at the single-molecule length scale[42] not just at an controlled, reductionist *in vitro* level[43] but also to gain molecular-level insight into the physiologically relevant context of single functioning, living cells.[44–49] Combining this automated CoPro method with multicolour single-molecule real-time fluorescence imaging[50,51] may also enable quantitative estimation of dynamic protein concentration changes of multiple interacting proteins in live cells, which is an appealing route towards investigating native biochemistry, one molecule at a time.

**Acknowledgements**

We thank Sviatlana Shashkova and Stefan Hohmann (University of Gothenburg, Sweden) for donation of yeast cell strains and assistance with yeast cell culturing. We thank Aisha Syeda and Peter McGlynn (University of York, UK) for assistance with adapting the DnaQ-YPet strain to DnaQ-GFP. With thank Holly Hathrell (University of York, UK) for assistance with characterizing the sensitivity of our spot detection algorithms. MCL is assisted by a Royal Society URF and research funds from the Biological Physical Sciences Institute (BPSI) of the University of York, UK.

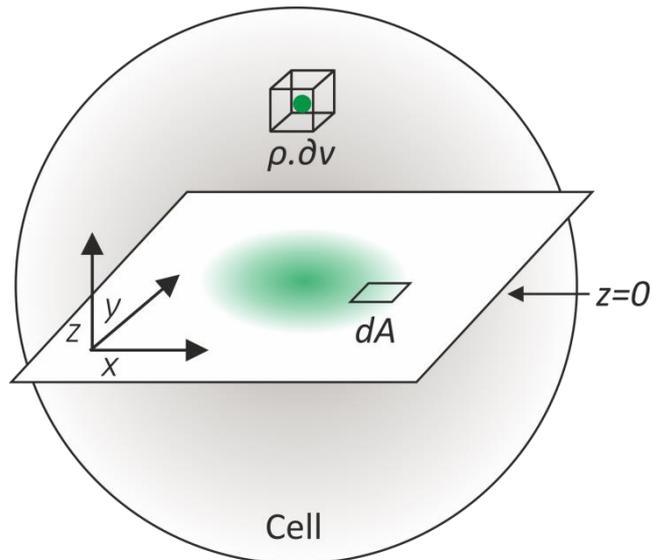

Figure 1: Diagram illustrating convolution integral over a single cell, in this simple example, assumed to be spherical. The intensity at an area element, *dA*, in the microscope's focal plane is the sum of the PSFs of all the fluorophores in the cell or the integral of the PSF multiplied by the concentration, *ρ*, over the cell volume, *V*. The focal plane is marked as *z=0*.

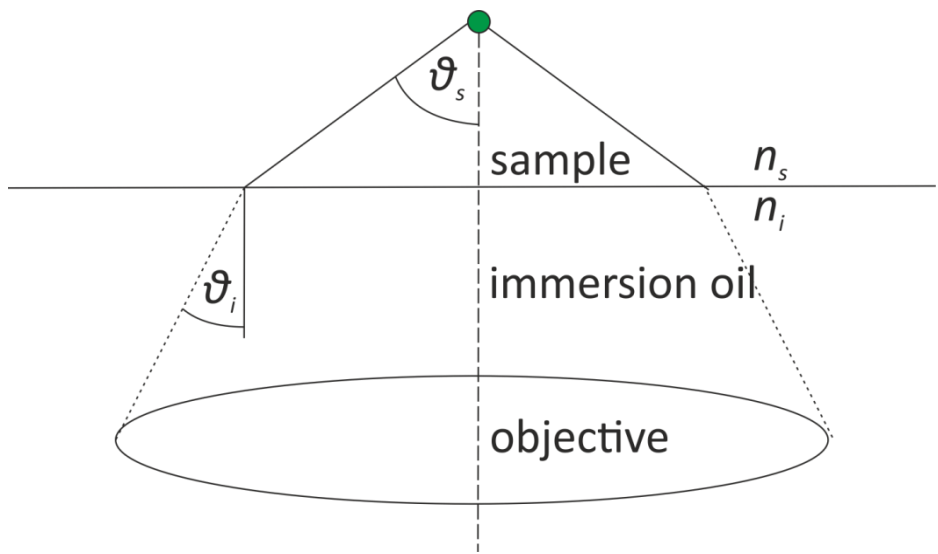

Figure 2: Schematic of light emitted by a point source collected by an objective lens. A ray of fluorescence emission light is traced from the source (green) at an angle $\vartheta_s$ in the sample media (i.e. water-based minimal media) with refractive index $n_s$, refracted at the interface of the immersion oil to an angle $\vartheta_i$ in the immersion oil media with refractive index $n_i$.

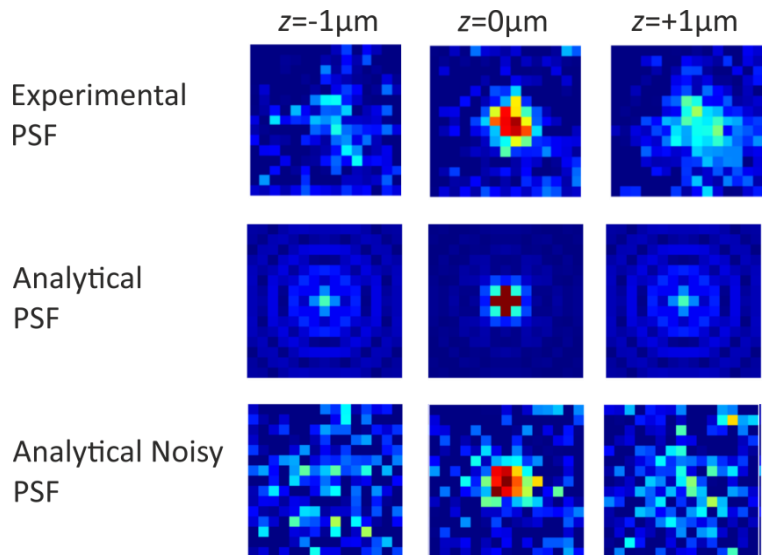

Figure 3: The measured, experimental PSF is shown (upper panel) over a 2 µm range in *z* centred on the focal plane (*z*=0), alongside the analytical PSF (middle panel) and the analytical PSF convolved with localisation error and noise for qualitative comparison (lower panel). The *chi*-squared value in comparing the analytical PSF model to the experimental over the *z* range -1 to +1 µm was 65 equating to a goodness of fit of probability confidence P<0.001.

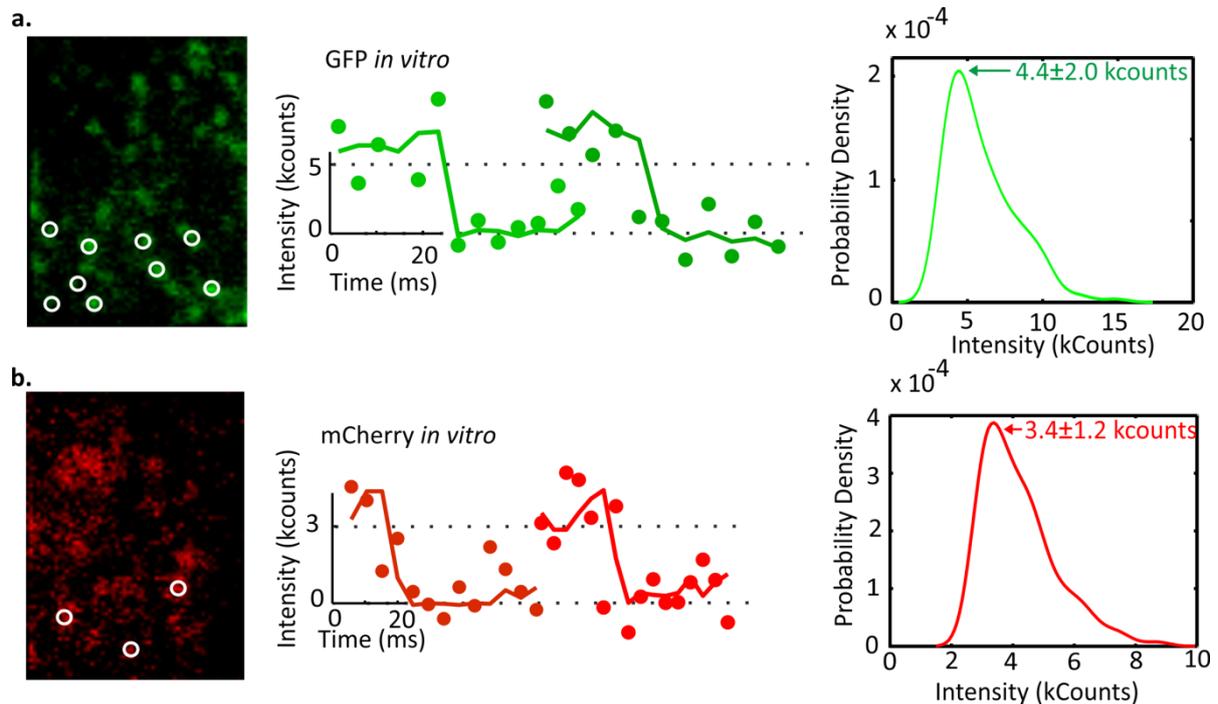

Figure 4: *In vitro* single-molecule fluorescent protein characterization. Surface immobilisation assay for **a.** GFP, and **b.** mCherry, showing typical fluorescence images (left panel, white indicating example autodetected spots from our bespoke localisation and tracking software); typical measured photobleach traces (middle panel); and probability distributions of single-molecule intensity values value (right panel) with peak and half width at half maximum (HWHM) error indicated.

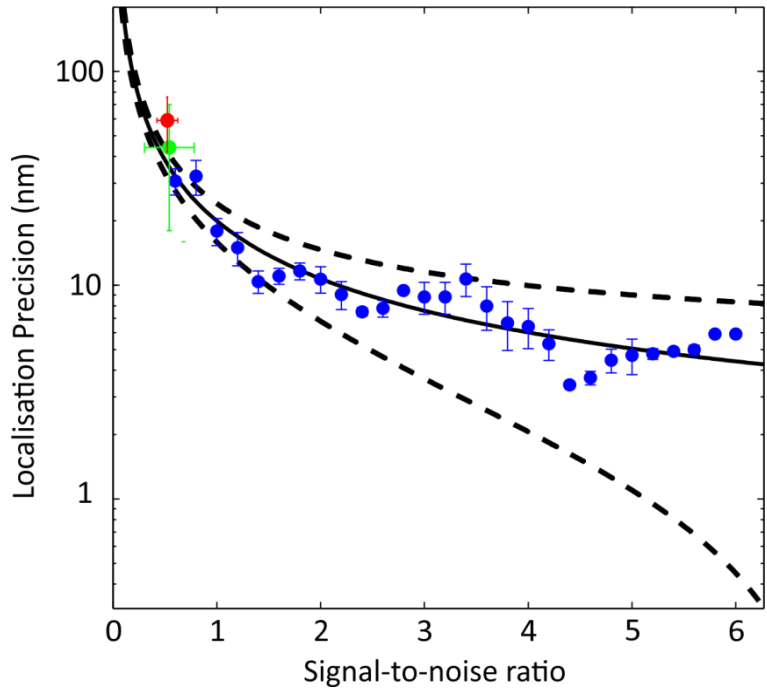

Figure 5: Mean localisation precision which we measured against signal-to-noise ratio for fluorescent beads (blue), GFP (green) and mCherry (red) with error bars showing the standard deviation on a semilog plot, each datapoint sampled from a set of *n*=5-10 beads. The black line is the fitted Thompson model for localisation precision (using parameters *b*=5, *G*=0.1) with 90% confidence bounds for fit (black dotted lines). The mean mCherry signal is smaller than GFP by ~26%, but also its PSF width is larger by ~10% resulting in a slight deviation from the black fit as predicated by the Thompson et al model.[36]

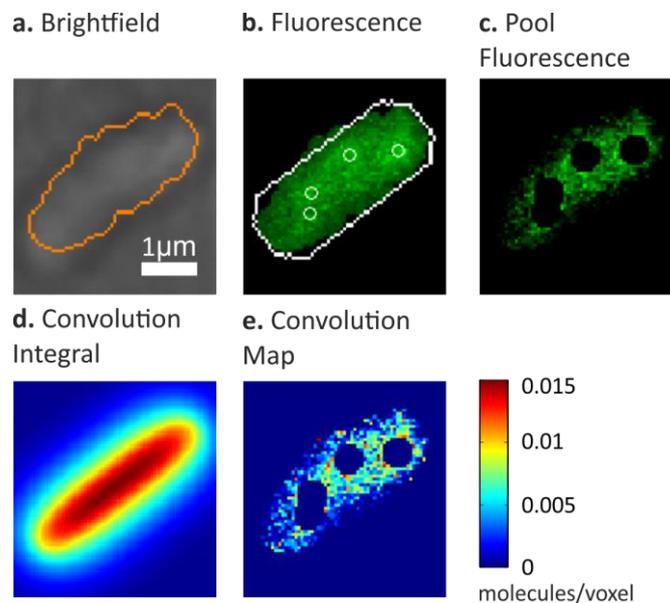

Figure 6: **a** Brightfield image (grey) of typical single live *E. coli* cell, with the segmented cell outline obtained from the fluorescence data (as outlined in the main text) shown overlaid (orange). **b** GFP fluorescence image (green) with found spots marked (white circles), white line is unique fitted 'sausage' function for that cell. **c** Pool fluorescent pixels with detected distinct spots now excluded from image. **d** Projection image of PSF corresponding to the *E.*

*coli* cell integrated over its cell volume. **e** Protein concentration map of cell (which is the image of **c** divided by image of **d** pixel-by-pixel).

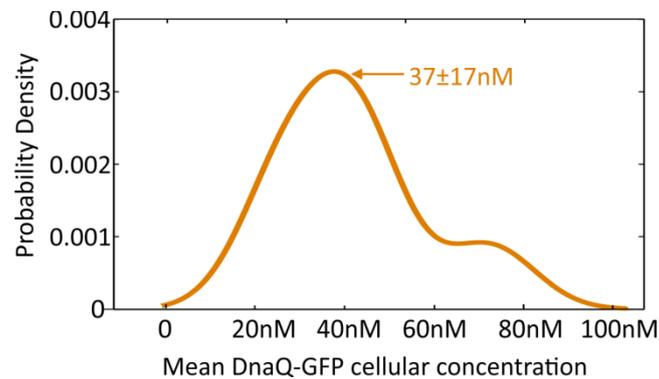

Figure 7: Example distribution of mean cellular DnaQ concentration which illustrates that even with only a few cells in a population (here, *n*=7 *E. coli* bacteria cells) we can reconstruct a sample probability distribution; mean and HWHM error indicated. The relative scale on the probability density axis is set to ensure normalization conditions (i.e. the area under the curve in exactly unity).

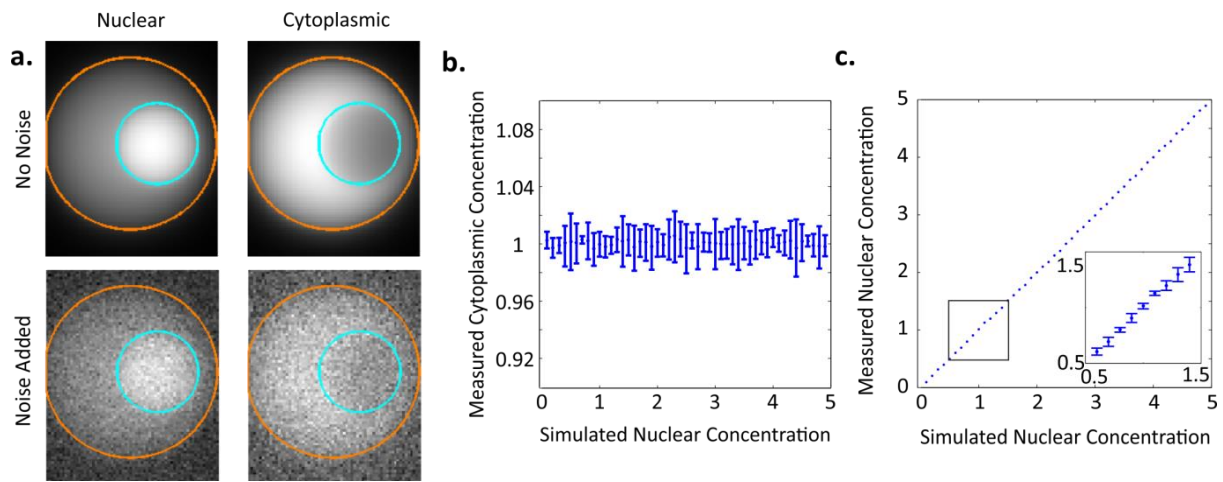

Figure 8: **a.** Simulated images of a spherical yeast cell with spherical nucleus, showing (left panel) nuclear concentration set to be twice the cytoplasmic concentration, and (right panel) zero concentration in the nucleus with all Mig1 localized in the cytoplasm. No noise (upper panel) and realistic noise added (lower panel) are shown. **b.** Measured cytoplasmic concentration for cells (*n*=300 cells, made up from 5 repeats at each of 60 different protein concentrations) with varying simulated nuclear concentration, standard deviation errorbars indicated. **c.** Measured nuclear concentration as a function of simulated nuclear concentration, with (inset) a zoom-in of the simulated data and the fit.

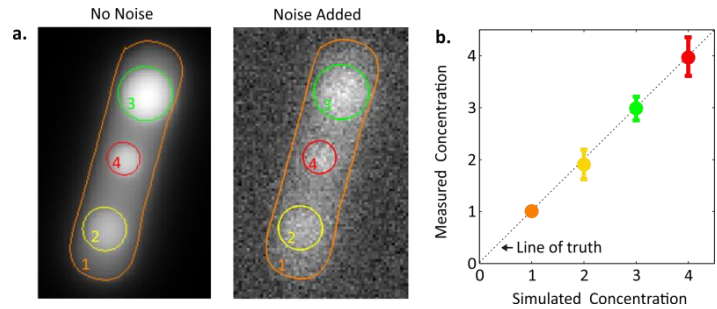

Figure 9: **a.** Simulated bacterial 'sausage' shaped cell (region 1) with three different sized spherical organelle compartments and different concentrations (regions 2-4). **b.** mean and standard deviation concentrations of these different regions measured using the CoPro method plotted against the real simulated values, with the 'line of truth' added (black dashed line).

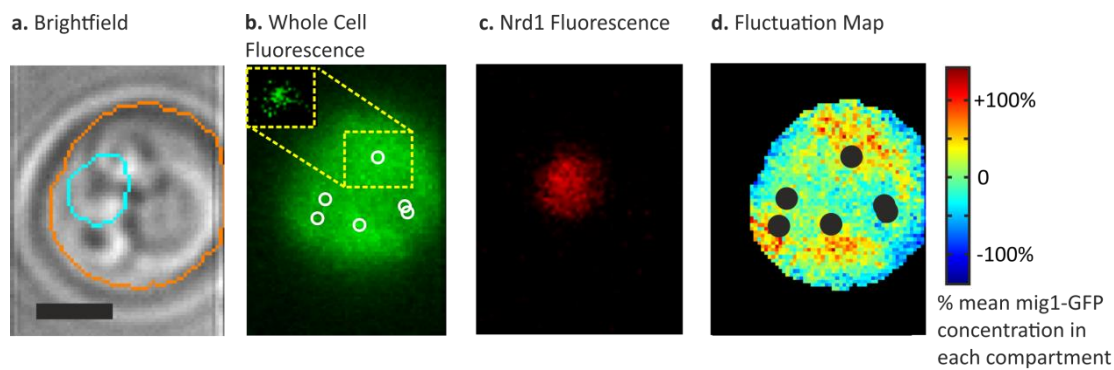

Figure 10: **a** Brightfield image of cell with automated segmentation outline of nucleus (cyan) and cell outer boundary (orange), 2 μm scale bar indicated. **b** GFP fluorescence image, reporting the localization of Mig1, with found spots (white circles) marked and cutaway (yellow) with adjusted intensity levels to indicate the position of the underlying detected spot. **c** mCherry fluorescence image, reporting the localization of Nrd1 in the nucleus. **d** Spatial distribution map of Mig1 concentration, indicating the fluctuation from the mean compartment concentration value with respect to position across the cell image.

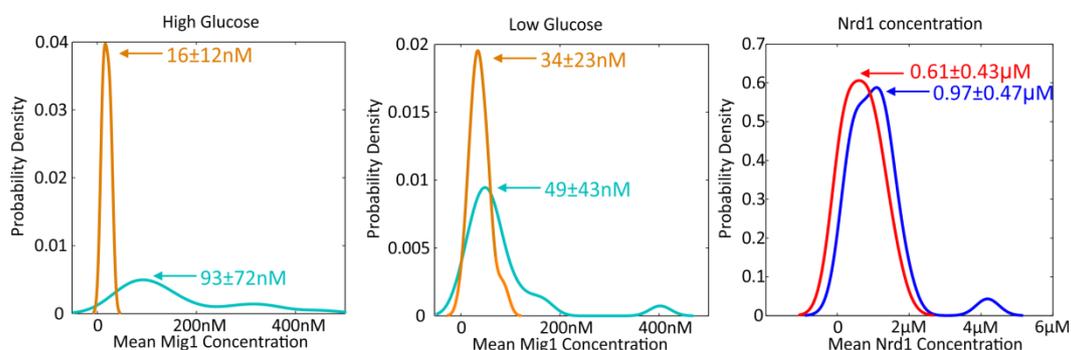

Figure 11: Distribution of cytoplasmic (orange) and nuclear (cyan) Mig1 protein concentrations of between *n*=25-30 cells at (a) high and (b) low glucose and distribution, and equivalent for (c) Nrd1 concentration (high glucose in blue and low glucose in red) with peak values ± HWHM indicated.